\documentclass[fleqn,usenatbib]{mnras}

\usepackage{newtxtext}
\usepackage{newtxmath}

\usepackage{natbib}
\usepackage[T1]{fontenc}
\usepackage{aecompl}

\usepackage{graphicx}							
\usepackage{bm}									
\graphicspath{{../graphics/}}					
\usepackage{amsmath}							
\usepackage{amssymb}							
\usepackage{physics}							
\usepackage[nameinlink, capitalize]{cleveref}	
\usepackage{array}

\newcommand{\erad}{\theta_\mathrm{E}}		
\newcommand{\brad}{\theta_\mathrm{B}}		
\newcommand{\trad}{\theta_\mathrm{T}}		
\newcommand{\critd}{\Sigma_\mathrm{c}}	
\newcommand{\reff}{r_\mathrm{eff.}}			
\newcommand{\etheta}{\theta_\varepsilon}	
\newcommand{\iu}{\mathrm{i}\mkern1mu}		
\newcommand{\kbr}{\kappa_\mathrm{B}}		
\newcommand{\ktr}{\kappa_\mathrm{T}}	

\title[Mass profiles from lensing III]{Galaxy mass profiles from strong lensing III: The two-dimensional broken power-law model}

\author[C. M. O'Riordan et al.]{
	C. M. O'Riordan,$^{1,2}$\thanks{E-mail: conor@mpa-garching.mpg.de}
	S. J. Warren,$^{1}$
	D. J. Mortlock$^{1,3,4}$
	\\
	$^{1}$Astrophysics Group, Blackett Laboratory, Imperial College London, London, SW7 2AZ, United Kingdom\\
	$^{2}$Max-Planck Institute for Astrophysics, Karl-Schwarzschild Str. 1, D-85748, Garching, Germany\\
	$^{3}$Department of Mathematics, Imperial College London, London, SW7 2AZ, UK\\
	$^{4}$Department of Astronomy, Stockholm University, Albanova, SE-10691 Stockholm, Sweden
}
\date{Accepted XXX. Received YYY; in original form ZZZ}
\pubyear{2020}

\begin{document}
\label{firstpage}
\pagerange{\pageref{firstpage}--\pageref{lastpage}}
\maketitle

\begin{abstract}
When modelling strong gravitational lenses, i.e., where there are multiple images of the same source, the most widely used parameterisation for the mass profile in the lens galaxy is the singular power-law model $\rho(r)\propto r^{-\gamma}$. This model may be insufficiently flexible for very accurate work, for example measuring the Hubble constant based on time delays between multiple images. Here we derive the lensing properties \---\ deflection angle, shear, and magnification \---\ of a more adaptable model where the projected mass surface density is parameterised as a continuous two-dimensional broken power-law (2DBPL). This elliptical 2DBPL model is characterised by power-law slopes $t_1$, $t_2$ either side of the break radius $\brad$. The key to the 2DBPL model is the derivation of the lensing properties of the truncated power law (TPL) model, where the surface density is a power law out to the truncation radius $\trad$ and zero beyond. This TPL model is also useful by itself. We create mock observations of lensing by a TPL profile where the images form outside the truncation radius, so there is no mass in the annulus covered by the images. We then show that the slope of the profile interior to the images may be accurately recovered for lenses of moderate ellipticity. This demonstrates that the widely-held notion that lensing measures the slope of the mass profile in the annulus of the images, and is insensitive to the mass distribution at radii interior to the images, is incorrect.
\end{abstract}

\begin{keywords}
	gravitational lensing: strong, galaxies: general
\end{keywords}

\defcitealias{ORiordan2019}{Paper I}
\defcitealias{ORiordan2020}{Paper II}
\defcitealias{Tessore2015a}{TM15}


\section{Introduction}
\label{sec:introduction}

This is the third in a series of papers examining how strong gravitational lensing of an extended source may be used to measure the
profile of the projected surface mass density in the lens, using lensing data alone, independent of dynamics. In the first and second papers in the series \citep[][hereinafter \citetalias{ORiordan2019} and \citetalias{ORiordan2020}]{ORiordan2019,ORiordan2020} we examined respectively the circular and elliptical cases of the singular power-law (SPL) lens, with 3D density profile $\rho(r)\propto r^{-\gamma}$.  The model projects to a power law in surface mass density with exponent $t=\gamma-1$. We used synthetic observations to determine the constraints on $\gamma$ for a wide range of image configurations, and we developed a theoretical understanding of how the different observables, positions and fluxes, contribute to the constraints on $\gamma$ in the various configurations. We showed that in the best cases the slope may be measured to an accuracy $\sigma_\gamma\simeq 0.01$.

Having shown that strong lensing observations on their own can provide accurate measurements of the slope we now turn to the question of where in the radial profile these constraints apply. In the SPL model the shape of the surface density profile is
quantified by a single parameter, the power-law exponent $\gamma$, or equivalently $t$. In the current paper we introduce a more versatile mass model with the goal of determining the constraints on the lens surface density as a function of projected radius. We define an elliptical model in which the surface density is a continuous broken power law (BPL). This 2DBPL model has two extra parameters compared to the SPL model: it is specified by an inner power-law slope $t_1$, an outer power-law slope $t_2$, and a break (elliptical) radius $\brad$. In the current paper we derive the main useful lensing properties of this model i.e. the deflection angle and the shear, and thereby the magnification. In the next paper in the series we will explore the constraints on the parameters $t_1, t_2, \brad$ for different image configurations.

The 2DBPL model, parameterising the surface density, is closely related to the 3DBPL model, parameterising the density, for which the lensing properties have been derived by \citet{Du2020}. The 3D model could be a physical model for a real galaxy, while the 2D model cannot be exactly, since no 3D ellipsoidal density distribution can project to a 2DBPL profile. The projection of the 3DBPL model to 2D results in a softened break between the two power laws. The original reason for developing the 2DBPL model was to have a distinct change in slope as a means of directly contrasting the model either 
side of the break. In this way by comparing the uncertainties on the two slopes it will be possible to gain an understanding of how lensing constrains the mass profile in the most direct manner.

Although created as a theoretical tool to understand how lensing constrains the mass profile, the 2DBPL profile is also
useful as a functional form for fitting real lenses as it represents the simplest extension beyond the commonly used SPL model. By using
Bayesian model comparison techniques it will be possible to determine if such an extension beyond the SPL model, with two extra parameters, is justified by the data. An accurate mass model is a critical component of any lensing study but this is especially true for time delay cosmography, where the delays in light travel time between the multiple images in a single system are used to constrain the Hubble constant \citep[e.g.][]{Wong2020}. A number of authors have suggested that the currently employed SPL model lacks the freedom necessary to model accurately the lens mass when measuring $H_0$ \citep{Xu2016,Sonnenfeld2018,Kochanek2020}. The consequence could be that measurements of $H_0$, although precise, are inaccurate because of a biased measurement of the potential at the positions of the lensed images. 

The lensing properties of the 2DBPL profile are derived by first determining the lensing properties of the 2D truncated power-law
(2DTPL) profile, where the surface density is represented by a singular power-law profile out to a specified radius, and
then zero beyond. The 2DBPL profile is then constructed by combining SPL and 2DTPL profiles in a straightforward way (explained
below). The derivation of the lensing properties of the 2DTPL model is therefore the main theoretical innovation presented in this paper.

For the remainder of the paper we refer to the 2DTPL and 2DBPL models simply as the TPL and BPL models for the sake of readability.

Regarding measurements of the mass profile, it is widely stated or implied that
lensing images constrain the slope of the profile only over the annulus spanned by the images, i.e. near the Einstein radius, and
provide no useful information on the mass profile interior to the innermost image \citep{Chae2014,Hezaveh2016,Kochanek1995,Kochanek2006,Koopmans2006,Spingola2018,Suyu2017,Treu2010a,vandeVen2009}. The clearest example of this sentiment is in \citet{Kochanek2006} where it is stated `it is important to remember that the actual constraints on the density structure really only apply over the range of radii spanned by the lensed images'. In the current paper we use the TPL model to examine this belief. We construct mock observations for elliptical lenses where the images form outside the truncation radius. In these observations there is therefore no mass in the annulus spanned by the lensed images. We then fit the TPL model to the mock observations to determine the constraints on the power-law slope for different image configurations. 

 The paper is organised as follows. In \cref{sec:truncated-power-law} we present the theory for the TPL  model. In \cref{sec:broken-power-law} we formulate the BPL  model and in \cref{sec:computation} we provide a numerical recipe for fast computation of the deflection angle for the TPL and BPL models. In \cref{sec:truncated-results} we use the TPL model to examine if the slope interior to the images can be constrained. Conclusions are presented in \cref{sec:summary}.

\section{Truncated Power-Law Mass Model}
\label{sec:truncated-power-law}

Before deriving the lensing properties for the BPL we first derive them for a simpler model, the truncated power-law (TPL). These results will be useful in formulating the BPL in \cref{sec:broken-power-law}.

\subsection{Convergence}
In the TPL model the surface density is described by a power-law interior to some elliptical radius called the truncation radius and is zero outside this radius. Explicitly, the convergence in the TPL model is
\begin{equation}
	\label{eq:truncated-convergence}
	\kappa(\etheta)=\begin{cases}
	\ktr(\trad/\etheta)^{t},&\quad \mathrm{for}\,\,\etheta\leq\trad\\
	0, &\quad \mathrm{for}\,\,\etheta>\trad,
	\end{cases}
\end{equation}
where $\trad$ is the truncation radius, $\ktr$ is the convergence at $\trad$, and $t$ is the logarithmic slope. The elliptical radius $\etheta$ is defined
\begin{equation}
	\label{eq:elliptical-radius}
	\etheta^2 = q^2\theta_1^2+\theta_2^2,
\end{equation}
where $q$ is the axis ratio of the minor to major axes of the mass distribution's isodensity contours, and is related to ellipticity $\varepsilon$ by $q=1-\varepsilon$. 

\Cref{eq:truncated-convergence} also includes a normalisation $\ktr$ which we now define. We wish to relate $\ktr$ to the scale length $b$ such that $b$ retains its usual meaning from the single power-law model; that  the average density inside the elliptical radius $b$ is the critical density $\critd$. If $M(\theta)$ is the mass enclosed by a radius $\theta$ then at $\theta=b$ we require
\begin{equation}
\label{eq:scale-radius}
M(b)=\critd \pi b^2 D_\mathrm{d}^2/q,
\end{equation}
where $D_\mathrm{d}$ is the distance to the lens. In general, for an elliptical mass distribution the total mass enclosed by a radius $\etheta$ is
\begin{equation}
\label{eq:mass-enclosed}
M(\etheta) = \frac{\critd D_\mathrm{d}^2}{q} \int_{0}^{\etheta} \kappa\left(\etheta'\right)2\pi \etheta'\dd\etheta'.
\end{equation}
Using \cref{eq:truncated-convergence}, for a radius $\etheta>\trad$ the mass enclosed is a constant;
\begin{equation}
	\label{eq:total-mass}
	M\left(\etheta>\trad\right)= \frac{2\pi \critd D_\mathrm{d}^2\ktr\trad^2}{q(2-t)}.
\end{equation}
To find $\ktr$ we can then combine \cref{eq:total-mass,eq:scale-radius} to obtain
\begin{equation}
	\ktr=\begin{cases}\vspace{0.2cm}
	\displaystyle\frac{2-t}{2\nu^2},&\quad \mathrm{for}\,\,\nu\leq1,\\
	\displaystyle\frac{2-t}{2\nu^t},&\quad \mathrm{for}\,\,\nu\geq1,
	\end{cases}
\end{equation}
where $\nu=\trad/b$.

\subsection{Deflection angle}
\label{sec:deflection-angle}

To find the deflection angle for the TPL model we follow initially the same route used by \citet[][hereafter \citetalias{Tessore2015a}]{Tessore2015a} for the SPL model, and adopt the same notation where the complex image plane coordinate is $z=\theta_1+\iu\theta_2$ and the complex deflection angle is $\alpha=\alpha_1+\iu\alpha_2$. \citet{Bourassa1975} give the deflection angle for a general elliptical mass profile as
\begin{equation}
	\label{eq:deflection-angle-definition}
	\alpha^*(z)=\frac{2}{qz}\int_0^{\etheta(z)}\kappa(\theta)\theta\left(1-q'\frac{\theta^2}{z^2}\right)^{-1/2}\dd\theta,
\end{equation}
where $q'=(1-q^2)/q^2$. Inserting \cref{eq:truncated-convergence} we have
\begin{equation}
	\label{eq:deflection-angle-integral}
	\alpha^*(z)=\frac{2\ktr\trad^t}{qz}\int_{0}^{\theta'}\theta^{1-t}\left(1-q'\frac{\theta^2}{z^2}\right)^{-1/2}\dd\theta,
\end{equation}
with
\begin{equation}
	\theta'=\begin{cases}
	\etheta(z),\quad &\mathrm{for}\,\,\etheta(z)\leq\trad,\\
	\trad,\quad &\mathrm{for}\,\,\etheta(z)>\trad.\\
	\end{cases}
\end{equation}
First consider the deflection angle outside $\trad$ for a circular mass distribution, i.e., with $q=1$. \Cref{eq:deflection-angle-integral} is then easy to evaluate and gives
\begin{equation}
	\label{eq:circular-truncated-deflection-angle}
	\alpha^*(z) = \frac{b^2}{z}.
\end{equation}
This is the same as for a point mass with Einstein radius $\erad=b$. As expected, the deflection angle exterior to the truncation radius does not depend	 on the distribution of mass interior, via $t$ or otherwise, when the lens is circular.
\begin{figure}
	\includegraphics[width=\columnwidth]{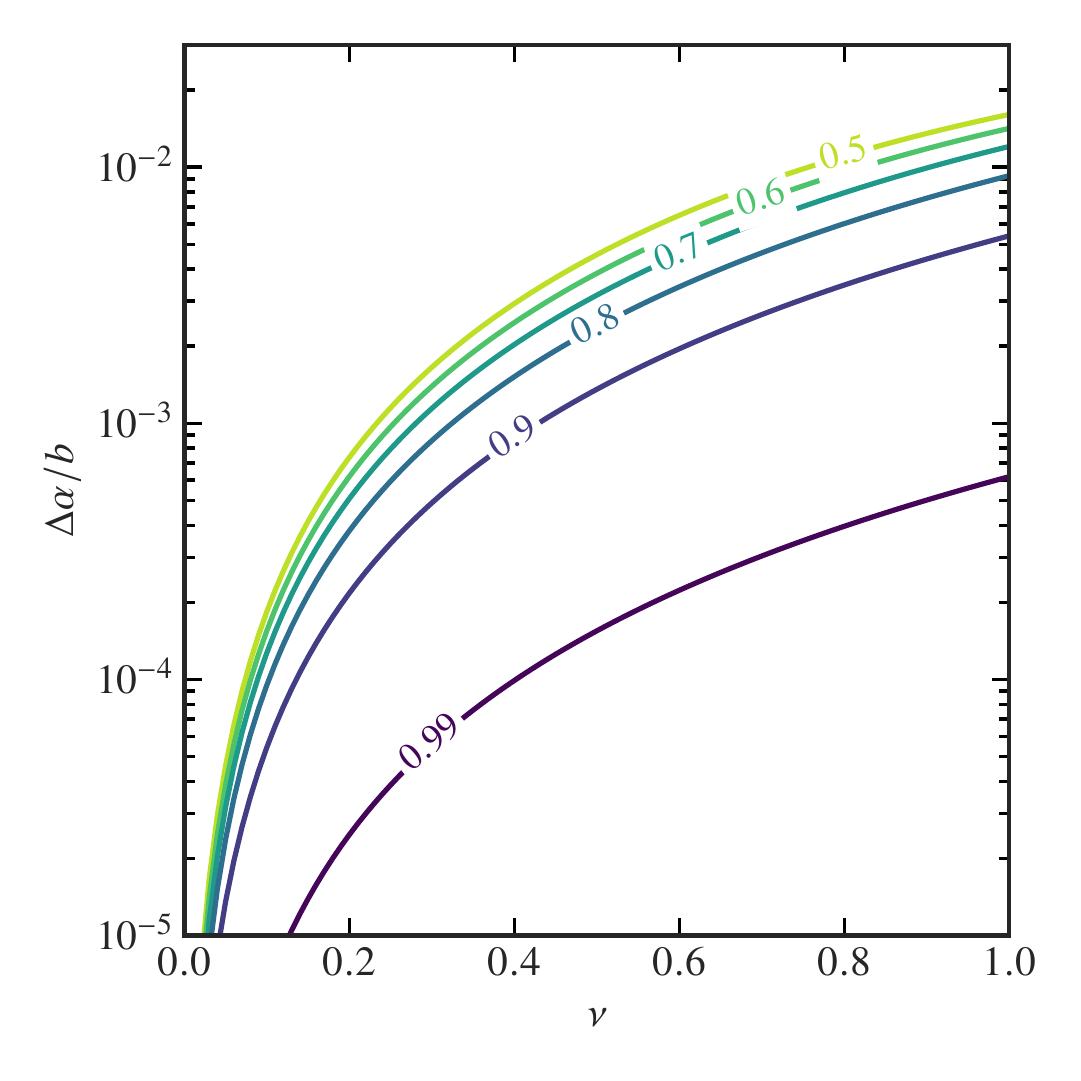}
	\caption{\label{fig:truncated-deflection-angle}
	The sensitivity of the deflection angle to small changes in slope as a function of $\nu=\trad/b$. The sensitivity is calculated along the line $\theta_1=\theta_2$ at the scale radius $\etheta=b$. Each curve represents a different axis ratio running from an (almost) circular lens to $q=0.5$.
}
\end{figure}

Now we consider the elliptical case, i.e., for a general $q\neq1$. Following \citetalias{Tessore2015a}, we take advantage of the fact that an integral of the form
\begin{equation}
	\label{eq:integral-definition}
	I(\theta,t;z)=\int_{0}^{\theta}\theta'^{1-t}\left(1-q'\frac{\theta'^2}{z^2}\right)^{-1/2}\dd\theta'
\end{equation}
has the solution
\begin{equation}
	I(\theta,t;z)=\frac{\theta^{2-t}}{2-t}F(\theta,t;z),
\end{equation}
where
\begin{equation}
	\label{eq:hypergeometric-definition}
	F(\theta,t;z)={}_2F_1\left(\frac{1}{2},1-\frac{t}{2};2-\frac{t}{2};q'\frac{\theta^2}{z^2}\right),
\end{equation}
and ${}_2F_1$ is the Gaussian hypergeometric function. \Cref{eq:deflection-angle-integral} then becomes
\begin{equation}
	\label{eq:deflection-angle-truncated}
	\alpha^*(z)=\frac{b^2}{qz}\times\begin{cases}
	\vspace*{0.2cm}
	\displaystyle F(\etheta,t;z)\left(\frac{\trad}{\etheta}\right)^{t-2} ,\quad &\mathrm{for}\,\,\etheta(z)\leq\trad,\\
	\displaystyle F(\trad,t;z),\quad &\mathrm{for}\,\,\etheta(z)\geq\trad.\\
	\end{cases}
\end{equation}
The deflection angle exterior to the truncation radius has now picked up a dependence on the mass distribution interior, through $t$ in the hypergeometric function. For the case $\etheta\leq \trad$ the deflection angle is identical to the final line of \citetalias{Tessore2015a}'s Eq. (11) up to a factor of $\nu^{2-t}$ which accounts for the mass deficit inside $b$ when $\nu<1$. This factor ensures that $b$ retains the same meaning for both the TPL and SPL even though they are differently normalised, i.e., a source at $\beta=0$ will form a ring at $\etheta=b$ when $q=1$ in both models. The result in the $\etheta(z)<\trad$ case could be further simplified by separating the radial and angular parts (see \citetalias{Tessore2015a}'s Eq. (12) onwards). In the $\etheta(z)>\trad$ case, $\alpha$ does not benefit from the same simplifications because the mass profile is only integrated up to $\trad$ in \cref{eq:deflection-angle-integral}, rather than all the way up to $z$. Note that the deflection angle is continuous across the boundary at $\trad$, even though the mass profile is not.

The sensitivity of the deflection angle outside $\trad$ to the slope inside $\trad$ is illustrated in \cref{fig:truncated-deflection-angle}. The quantity $\Delta \alpha$, as a fraction of $b$, is calculated at the point at elliptical radius $b$ on the diagonal line $\theta_1=\theta_2$. $\Delta \alpha$ is the magnitude of the change in the deflection angle vector for a change in slope of $\Delta t=0.1$ around an isothermal, i.e., $t=1$ slope. The figure shows how the sensitivity depends on two features of the lens. First, the deflection angle is more sensitive to the mass interior as the truncation radius approaches the Einstein radius, i.e., as $\nu=\trad/b\rightarrow1$. For a very small truncation radius the dependence vanishes entirely, and as $\nu\rightarrow1$ the quantity $\Delta \alpha$ in the figure converges to that of the SPL. Second, the dependence is stronger for higher ellipticities. This could have been predicted from the results of \citetalias{ORiordan2020} where we found that in general more elliptical lenses provide better constraints on the mass profile slope. The large increase in sensitivity between $q=0.99$, essentially a circular lens, and $q=0.9$ indicates that the images should be sensitive to the interior slope even for small ellipticities. We will examine both of these effects in more detail in the next paper in this series.

\subsection{Shear and magnification}
\label{sec:shear}

The complex shear $\gamma(z)$ is defined as
\begin{equation}
	\label{eq:shear-definition}
	\gamma^*(z) = \pdv{\alpha^*}{z},
\end{equation}
where 
\begin{equation}
	\pdv{z}=\frac{1}{2}\left(\pdv{\theta_1}-\iu\pdv{\theta_2}\right),
\end{equation}
is the Wirtinger derivative. The shear interior to the truncation radius is just that for the single power-law, given by \citetalias{Tessore2015a}'s Eq. (16) but with our definitions of $\kappa$ and $\alpha$. Exterior to the truncation radius we use the result
\begin{equation}
	\label{eq:hypergeometric-derivative}
	\pdv{z}F(\theta,t;z)=\frac{2-t}{z}\left[F(\theta,t;z)-\left(1-q'\frac{\theta^2}{z^2}\right)^{-1/2}\right],
\end{equation}
to find
\begin{equation}
	\label{eq:truncated-shear}
	\gamma^*(z)=\begin{cases}
		\vspace{0.1cm}
		\displaystyle(1-t)\frac{\alpha^*(z)}{z}-\frac{z^*}{z}\kappa(z) ,\quad &\mathrm{for}\,\,\etheta(z)\leq\trad,\\
		\displaystyle(1-t)\frac{\alpha^*(z)}{z}-\frac{2-t}{z^2}\left(1-q'\frac{\trad^2}{z^2}\right)^{-1/2} &\mathrm{for}\,\,\etheta(z)\geq\trad.\\
	\end{cases}
\end{equation}
Having defined both the convergence and shear it is possible to compute the scalar magnification from the standard relation
\begin{equation}
	\mu^{-1}=(1-\kappa)^2-\abs{\gamma}^2.
\end{equation}

\subsection{Lensing potential}
\label{sec:lensing-potential}
Inside the truncation radius the potential is simply that due to a single power-law which \citetalias{Tessore2015a} gives as
\begin{equation}
	\label{eq:truncated-potential-inside}
	\psi(z) = \frac{z\alpha^*(z)+z^*\alpha(z)}{2(2-t)},
\end{equation}
where $\alpha^*(z)$ is given by \cref{eq:deflection-angle-truncated}. The potential outside the truncation radius due to the mass interior can be found by summing the contributions from infinitesimally thin, homoeoidal, concentric elliptical rings each of constant density. From \citet{Schramm1990} we get the potential for such a ring at a position $z$ outside the ring
\begin{equation}
	\psi(z) = \frac{2}{q}\kappa(\etheta)\etheta \cosh[-1](\frac{zq}{\etheta\sqrt{1-q^2}})\dd{\etheta},
\end{equation}
where $q$ is the axis ratio of the ring, in this case constant across all rings, $\kappa$ is the surface density of the ring, and $\etheta$ is the elliptical radius of the ring defined in \cref{eq:elliptical-radius}. Using \cref{eq:truncated-convergence} we can then find the potential for the mass inside $\trad$ with the integral
\begin{equation}
	\label{eq:truncated-potential-outside}
	\psi(z) = \frac{2\ktr\trad^t}{q}\left[\int_{0}^{\trad}\theta^{1-t}\cosh^{-1}\left(\frac{zq}{\theta\sqrt{1-q^2}}\right)\dd{\theta}\right]  - C_\mathrm{T},
\end{equation}
which must be evaluated numerically. The constant $C_\mathrm{T}$ ensures the potential is continuous across $\trad$. Explicitly this is $C_\mathrm{T}=\psi_{\etheta>\trad}(z_\mathrm{T})-\psi_{\etheta<\trad}(z_\mathrm{T})$ where each $\psi$ is given by \cref{eq:truncated-potential-inside,eq:truncated-potential-outside} respectively and $z_\mathrm{T}$ is an arbitrary point at the truncation radius, e.g. $z_\mathrm{T}=\iu \trad$.

\section{Broken Power-Law Mass Model}
\label{sec:broken-power-law}
We can make use of the results for the TPL to formulate a more useful model, a two-dimensional continuous broken power-law (BPL). The BPL model that follows is implemented in the lens modelling software \href{https://github.com/Jammy2211/PyAutoLens}{PyAutoLens} \citep{Nightingale2018}.

\subsection{Convergence}
We define the convergence for a BPL projected mass-density profile as
\begin{equation}
\label{eq:bpl-convergence}
\kappa(\etheta) = \begin{cases}\displaystyle\kbr(\brad/\etheta)^{t_1},&\mathrm{for}\,\,\etheta\leq\brad,\\
\displaystyle\kbr(\brad/\etheta)^{t_2},&\mathrm{for}\,\,\etheta\geq\brad.
\end{cases}
\end{equation}
where $\brad$ is the break radius, $\kbr$ is the convergence at the break radius, and $t_1$ and $t_2$ are the inner and outer 2D logarithmic slopes respectively. Inserting \cref{eq:bpl-convergence} into \cref{eq:mass-enclosed} gives an expression for $\kbr$,
\begin{equation}
\label{eq:bpl-normalisation}
\kbr=\begin{cases}
{\displaystyle\frac{2-t_1}{2\nu^2 \left[1+\delta_t \left(\nu^{t_2-2}-1\right)\right]}},&\mathrm{for}\,\,\nu<1,\vspace{0.2cm}\\
{\displaystyle\frac{2-t_1}{2\nu^{t_1}}},&\mathrm{for}\,\,\nu\geq 1,
\end{cases}
\end{equation}
where $\nu$ now has the definition $\nu=\brad/b$ and we define
\begin{equation}
\delta_t = \frac{2-t_1}{2-t_2}.
\end{equation}

\subsection{Deflection Angle}

We can find the deflection angle from \cref{eq:deflection-angle-definition}. The solution contains integrals of the form in \cref{eq:integral-definition} which we use to simplify the results.

For positions inside the break radius, i.e. where $\etheta\leq\brad$, we find
\begin{equation}
\label{eq:deflection-angle-bpl-interior}
\alpha^*(z)=\frac{2\kbr}{2-t_1}\frac{\brad^2}{qz}\left(\frac{\brad}{\etheta}\right)^{t_1-2} F(\etheta,t_1;z),
\end{equation}
which is just the deflection angle for a power-law of slope $t_1$ with an adjusted normalisation. Outside the break radius, i.e. where $\etheta\geq\brad$, we have
\begin{align}
\label{eq:deflection-angle-bpl-exterior}
\alpha^*(z)=\frac{2\kbr}{2-t_1}\Biggl\{&\frac{\brad^2}{qz}F(\brad,t_1;z)\\
+\delta_t\Biggl[&\frac{\brad^2}{qz}F(\etheta,t_2;z)\left(\frac{\brad}{\etheta}\right)^{t_2-2} - \frac{\brad^2}{qz}F(\brad,t_2;z)\Biggr]\Biggr\}\nonumber,
\end{align}
which can be further simplified if $\nu<1$ using \cref{eq:bpl-normalisation} for $\kbr$. Writing the deflection angle in this way is useful because it makes its three constituent parts clear. Comparing with \cref{eq:deflection-angle-truncated} we see that $\alpha(z)$ for the BPL can be found by combining three simpler deflection angles. We add $\alpha$ for a power-law with the inner slope $t_1$ truncated at $\brad$ and $\alpha$ for a single power-law with the outer slope $t_2$ (first and second terms). We then subtract the contribution from the single power-law with slope $t_2$ inside the break radius (third term).

We can check the result for the case where both slopes are the same, i.e., when $t_1=t_2=t$. In this case the normalisation becomes
\begin{equation}
\kbr = \frac{2-t}{2\nu^t	}.
\end{equation}
We also have $\delta_t=1$ and so the first and third hypergeometric terms in \cref{eq:deflection-angle-bpl-exterior} cancel out. Both \cref{eq:deflection-angle-bpl-interior,eq:deflection-angle-bpl-exterior} reduce to
\begin{equation}
\label{eq:deflection-angle-spl}
\alpha^*(z) = \frac{b^2}{qz}\left(\frac{b}{\etheta}\right)^{t-2}F(\etheta,t;z),
\end{equation}
which is identical to \citetalias{Tessore2015a}'s Eq. (11), i.e., the deflection angle for a single power-law.

\subsection{Shear}

We find the shear for the BPL by using \cref{eq:deflection-angle-bpl-interior,eq:deflection-angle-bpl-exterior} with the definition of complex shear in \cref{eq:shear-definition}. For the case where $\etheta<\brad$ we have
\begin{equation}
\gamma^*(z)=\frac{2\kbr \nu^{t_1}}{2-t_1}\,\gamma^*_{t_1}(z),
\end{equation}
where $\gamma^*_{t}(z)$ is the shear due to a single power-law with slope $t$, given by
\begin{equation}
\gamma^*_{t}(z) = (1-t)\frac{\alpha_t^*(z)}{z}-\kappa(z)\frac{z^*}{z},
\end{equation}
with $\alpha_t^*(z)$ given by \cref{eq:deflection-angle-spl}. Outside the break radius, when $\etheta>\brad$, the shear is
\begin{align}
\gamma^*(z)=\frac{2\kbr}{2-t_1}\Biggl\{\frac{\brad^2}{qz^2}\Biggl[&\frac{t_1-t_2}{\sqrt{1-q'\brad^2/z^2}}+\,(1-t_1)F(\brad,t_1;z)\nonumber\\&-\,(1-t_2)F(\brad,t_2;z)\Biggr]
+\,\nu^{t_2}\delta_t\gamma_{t_2}^*(z)\Biggr\}.
\end{align}
Comparing with the shear for the truncated profile in \cref{eq:truncated-shear}, we see that the shear for the BPL has the same composition as described for the deflection angle in the previous section.

\section{Computation}
\label{sec:computation}

To perform MCMC fitting to image planes using either the TPL or BPL we need a method to efficiently compute the deflection angles in \cref{eq:deflection-angle-truncated,eq:deflection-angle-bpl-interior,eq:deflection-angle-bpl-exterior}. In the case of both models, all that is required is a method for computing our specific hypergeometric function $F$ in \cref{eq:hypergeometric-definition}. We pursue a similar approach to that of \citetalias{Tessore2015a} for the SPL, however in this work we cannot simplify the final argument of $F$ in the same way and so we require a slightly different recipe for computation. In \citetalias{Tessore2015a} the $\theta$ used in $F$ is simply $\etheta(z)$ which allows for some simplification using their definition of $z=\etheta\mathrm{e}^{\iu \phi}$, where $\phi$ is an elliptical angle. In this work we must be able to use the constants $\brad$ or $\trad$ in place of $\etheta(z)$, prohibiting the same simplifications.

Recall that we defined
\begin{equation}
F(\theta,t;z)={}_2F_1\left(\frac{1}{2},1-\frac{t}{2};2-\frac{t}{2};q'\frac{\theta^2}{z^2}\right).
\end{equation}
We can exploit the fact that $F$'s parameters are of the form $a+b'+\frac{1}{2}=c$, allowing us to use the quadratic transformation
\begin{equation}
{}_2F_1\left(a,b';a+b'+\frac{1}{2},z'\right)={}_2F_1\left(2a,2b';a+b'+\frac{1}{2},\frac{1-\sqrt{1-z'}}{2}\right),
\end{equation}
\citep{Bateman1995}. Under this transformation $F$ becomes
\begin{equation}
F(\theta,t;z)={}_2F_1\left[1,2-t;2-\frac{t}{2};u(\theta;z)\right],
\end{equation}
where
\begin{equation}
u(\theta;z)=\frac{1-\sqrt{1-q'\theta^2/z^2}}{2}.
\end{equation}
With the condition that $\abs{u}<1$ we can use the series representation of ${}_2F_1$, as follows
\begin{equation}
\label{eq:expasion}
F(\theta,t;z)=\sum_{n=0}^{\infty}a_n(t) \,u(\theta;z)^n
\end{equation}
where
\begin{equation}
a_n(t) = \frac{\Gamma(n + 1)}{n!} 
\frac{\Gamma(2 - t + n)}{\Gamma\left(2- t/2 + n\right)}
\frac{\Gamma\left(2-t/2\right)}{\Gamma(2 - t)}.
\end{equation}
The quadratic transformation ensures that $\abs{u}$ only exceeds unity for very extreme axis ratios. For axis ratios $q\lesssim0.32$ there are regions in the image plane where $\abs{u}>1$ and the method does not produce accurate results. For our purposes however, the range of axis ratios $0.32<q\leq1$ is more than sufficient.

By using the fact that $\Gamma(z+1)=z\Gamma(z)$ we find a relatively simple recurrence relation between consecutive coefficients
\begin{equation}
\frac{a_{n+1}}{a_n} = \frac{2n + 4 - 2t}{2n + 4 - t},
\end{equation}
and $a_0(t)=1$. This leaves us with a straightforward recipe for computing $\alpha$ in the BPL with
\begin{equation}
\alpha^*(z)=\frac{2\kbr}{qz}\frac{\brad^2}{2-t_1}\left(\frac{\brad}{\etheta}\right)^{t_1-2}\sum_{n=0}^{\infty}a_n(t_1) \,u(\etheta;z)^n,
\end{equation}
for $\etheta\leq\brad,$ and
\begin{align}
\alpha^*(z)=\frac{2\kbr}{qz}\frac{\brad^2}{2-t_1}\sum_{n=0}^{\infty}\Biggl\{
&\,u(\brad;z)^n\left[a_n(t_1)-\delta_t a_n(t_2)\right]\\
+\,&u(\etheta;z)^n\left(\frac{\brad}{\etheta}\right)^{t_2-3}\delta_t a_n(t_2)\,\Biggr\},\nonumber
\end{align}
for $\etheta>\brad$.

\begin{figure}
	\includegraphics[width=1.0\columnwidth]{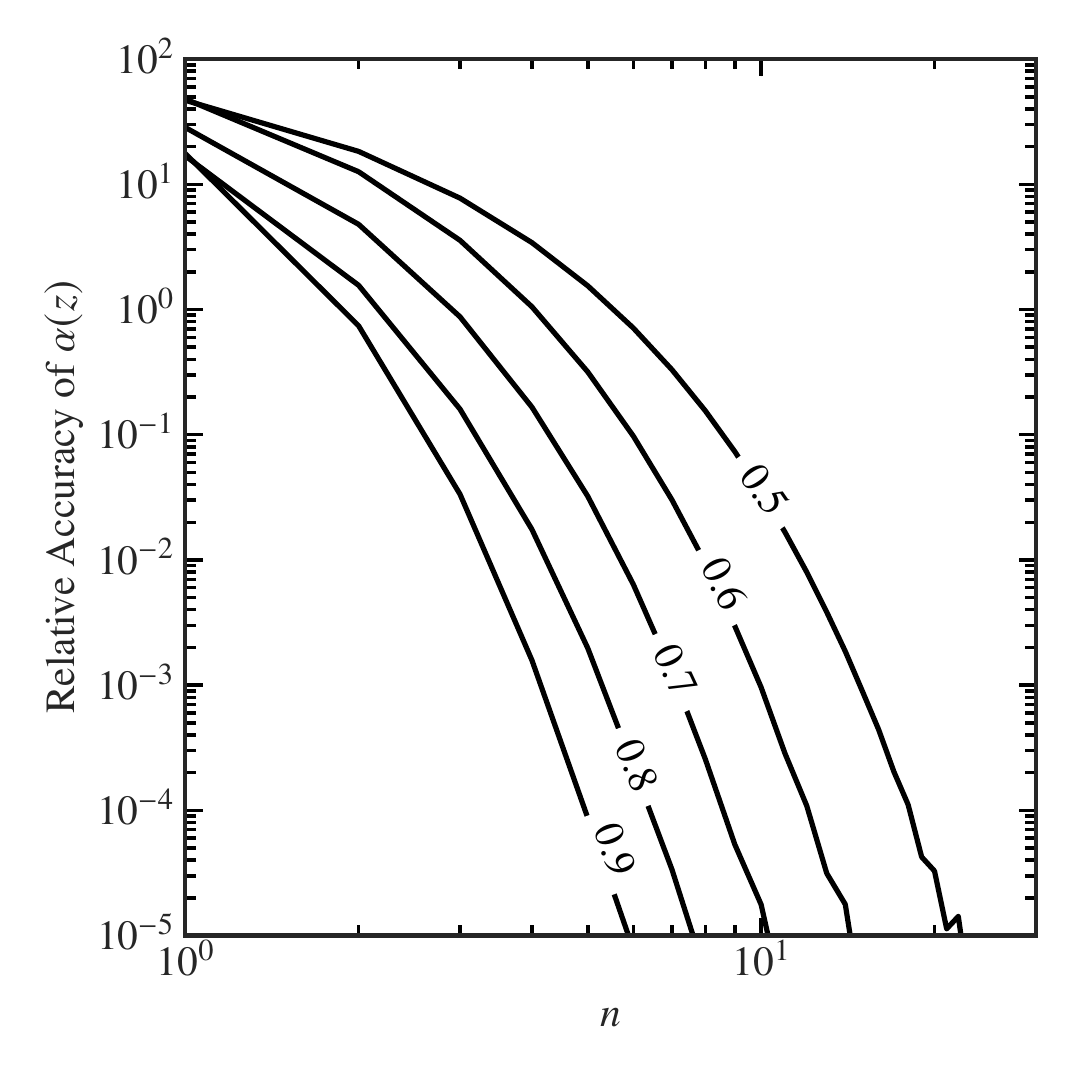}
	\caption{\label{fig:accuracy} The relative accuracy of \cref{eq:expasion} for different values of $n$. Each line corresponds to a different axis ratio, which is labelled.}
\end{figure}

\Cref{fig:accuracy} shows the relative accuracy of the method as the number of coefficients $n$ increases. In this case the relative accuracy is the maximum (over a $100\times100$ pixel 6 arcsec wide image plane) relative difference between the computed deflection angle using $n$ coefficients and the same with $n=100$. Computation time increases linearly with the number of coefficients.

\section{Constraining the slope interior to the images}
\label{sec:truncated-results}

We now return to the TPL model developed in \cref{sec:truncated-power-law} and its primary use in addressing a central question in this series of papers: what constraints, if any, can one obtain on the mass distribution interior to the lensed images? For a circular TPL model the deflection angle outside $\trad$ has no dependence on the mass profile interior to $\trad$ (see \cref{eq:circular-truncated-deflection-angle}). The lensing potential outside any circularly symmetric mass distribution is that of a point mass at the centre. However, this does not hold for elliptical mass distributions.

\begin{figure}
	\includegraphics[width=1.0\columnwidth]{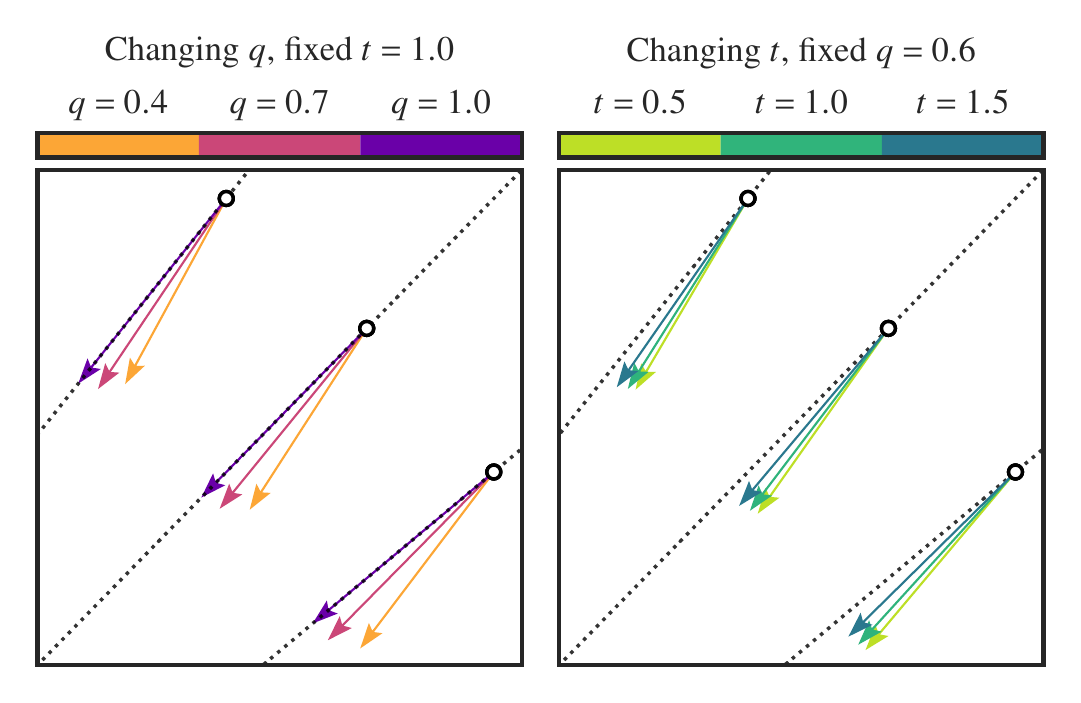}
	\caption{\label{fig:thought-experiment} The effect of the axis ratio $q$ and the slope $t$ on the deflection angle in a truncated lens. The vectors are the deflection angle (not to scale) at three points near the Einstein radius in the upper right quadrant of the image plane. The truncation radius is inside these points. The dotted lines intersect the centre of the lens.}
\end{figure}

As a thought experiment, consider the potential $\psi(z)$ and the deflection angle $\alpha(z)$ at a point $z$ outside the truncation radius, and not lying on a principal axis, as in \cref{fig:thought-experiment}. For a circular lens the vector points towards the centre of the lens. This is the case in the left panel of \cref{fig:thought-experiment} when $q=1$, $t=1$ and is obvious from \cref{eq:circular-truncated-deflection-angle}. As the ellipticity increases the deflection angle points away from the centre and towards the lens's semi-major axis, or the $x$-axis in the figure. Increasing (decreasing) the ellipticity decreases (increases) the circularity of the potential $\psi(z)$ and, because $\alpha=\pdv*{\psi}{z}$, the deflection angle changes direction accordingly.

Increasing the ellipticity, but keeping the same total mass inside $\trad$, moves mass from the centre to the region around the semi-major axis. In principle, a similar result can be achieved by changing the slope $t$. Increasing $t$ concentrates mass in the centre of the lens, circularising the potential and the deflection angle points more towards the centre as a result. Conversely, decreasing $t$ pushes mass towards the edge of the lens and the potential also becomes more elliptical. This behaviour is clear in the right-hand panel of \cref{fig:thought-experiment}.

This thought experiment raises two interesting points. First, for an elliptical lens with multiple images exterior to the truncation radius, the deflection angle at the images must be sensitive to the slope in the centre. Second, both the slope $t$ and the axis ratio $q$ can be used as we described to modify the angular structure in the potential. For small changes in either variable these modifications should look very similar. We therefore expect $t$ and $q$ to be correlated.

Rather than restrict ourselves to this thought experiment, we can perform an actual experiment using the TPL model we developed in \cref{sec:truncated-power-law}. We construct a number of mock observations with the technique used elsewhere in the series. The details of this procedure are in Section 3 of \citetalias{ORiordan2019}. The fidelity of the procedures for creating mock observations has been confirmed by the good agreement between parameters and their uncertainties derived from analysis of the mock observations and the predictions of theory \citepalias{ORiordan2019,ORiordan2020}.

\begin{figure*}
	\includegraphics[width=0.99\textwidth]{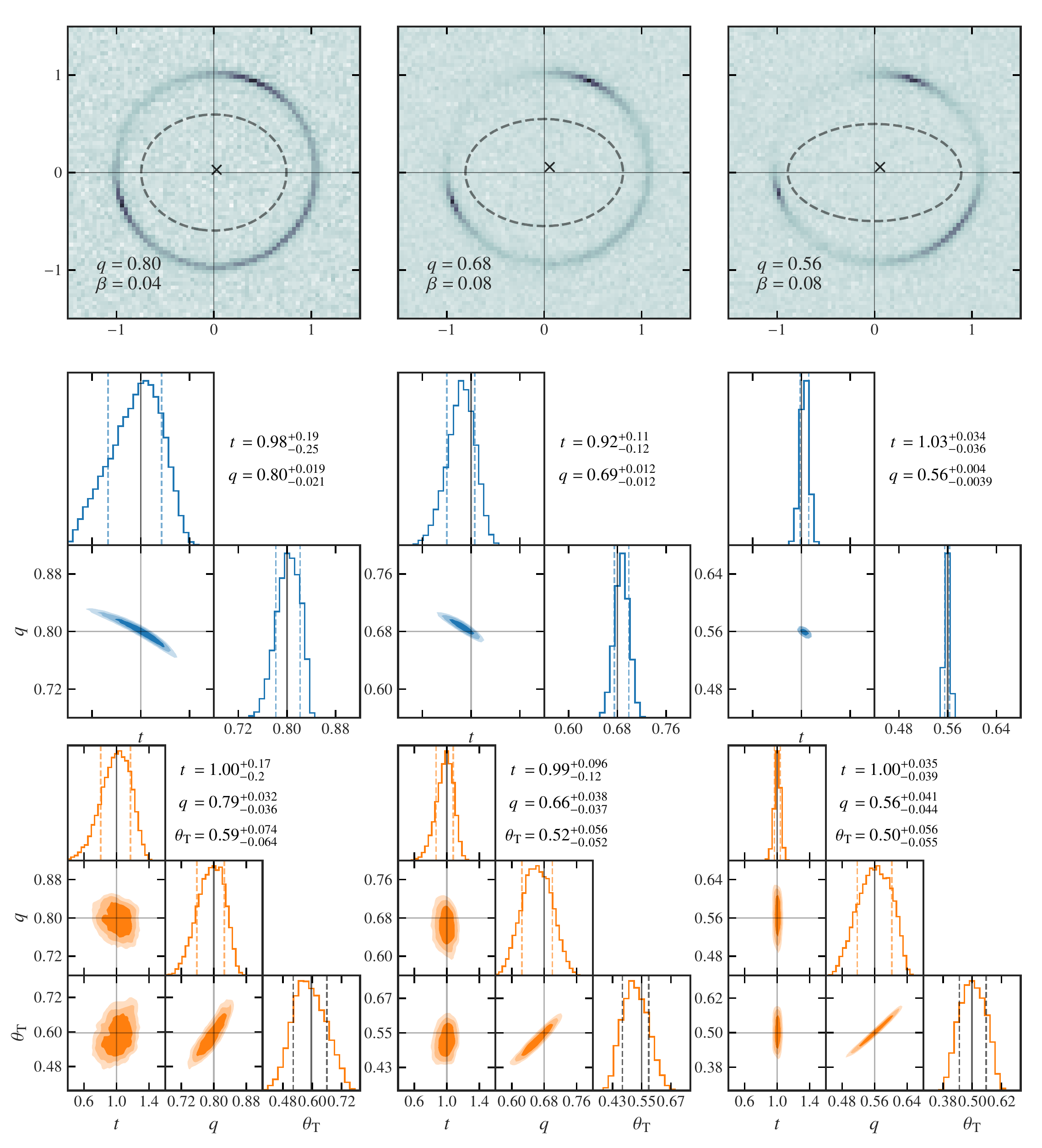}
	\caption{\label{fig:truncated-results} Constraints on the mass profile slope $t$, axis ratio $q$ and break radius $\trad$ for three truncated systems, created with signal to noise ratio $S=100$. {\em Top row:} Image plane. The dashed ellipse marks the truncation radius and the cross is the source position. {\em Middle row:} Case 1 results, $\trad$ fixed. {\em Bottom row:} Case 2 results, $\trad$ free. Contours are the 68\%, 95\%, and 99\% credible intervals. The 68\% credible intervals are also shown as dashed lines in the 1D histograms. Solid grey lines in the histograms mark the true values.}
\end{figure*}

We created mock observations with the TPL model, ensuring that in all cases the images form outside the truncation radius. This ensures that in the annulus spanned by the images there is no mass whatsoever, and likewise, in the region where there is mass, there are no images.\footnote{The infinite width of the S\'ersic profile means there is actually a vanishingly small amount of source flux inside the truncation radius. In the mock observation with an image closest to the truncation radius (the third observation in \cref{fig:truncated-results}) the ratio of source flux inside $\trad$ to the total flux is 0.002. In general the source flux interior to $\trad$ is well below the average noise level and will not contribute to the fit.} All mock observations have the same total signal to noise ratio in the images $S=100$. $S$ is defined such that only pixels which belong to lensed images, found by a masking procedure, are included in the calculation (see \citetalias[][\S 3.3]{ORiordan2019}). This means that the ability of different lens configurations to constrain parameters can be compared in an equal manner, independently of the magnification. Our mock observations use a similar set of parameters to those for the circular and elliptical SPLs in previous papers. We have $b=\sqrt{q}$, $t=1$ (equivalent to $\gamma=2$) and use a S\'ersic source with index $n_s=2$. 

Examples of mock observations are provided in the top row of \Cref{fig:truncated-results}, with $q$ decreasing, ellipticity increasing, left to right. The dashed ellipse marks the truncation radius in each case. The images produced by the TPL model are both significantly thinner and significantly more circular in nature compared to the SPL model with the same parameters. These images could be directly compared with, for example, Figure 2 of \citetalias{ORiordan2020}. The potential for the TPL model is both steeper than in the SPL and circularises faster. The result is a smaller radial magnification and a more `full' Einstein ring between the images, because the caustic is smaller.

We used ensemble Markov chain Monte Carlo (MCMC) sampling to fit two different models to the data. These both use the TPL profile to model the mass and a S\'ersic profile to model the source. In the first case, Case 1, the truncation radius is fixed to the true value. In Case 2 the truncation radius is a free parameter. Case 1 and Case 2 have 11 and 12 parameters respectively. For each case there are seven parameters for the source: position $\beta_x$ and $\beta_y$, size $\reff$, brightness $I_0$, S\'ersic index $n_\mathrm{s}$, axis ratio $q_\mathrm{S}$ and position angle $\phi_\mathrm{S}$. Both models have four lens parameters in common: lensing strength $b$, mass profile slope $t$, axis ratio $q$, and position angle $\phi_\mathrm{L}$. Case 2 has the extra parameter $\trad$.  

\begin{table}
	\centering
	\begin{tabular}{ c c c c }
		System & $\chi^2$ & $N_\mathrm{dof}$ & $\frac{\chi^2 - N_\mathrm{dof}}{\sqrt{2N_\mathrm{dof}}}$\\
		\hline
		1 & 533.76 & 573 & -1.16\\
		2 & 598.36 & 541 & 1.74\\
		3 & 529.89 & 485 & 1.44\\
		\hline
	\end{tabular}
	\caption{\label{table:chisq} Goodness-of-fit for the recovered parameter values in the three systems in \cref{fig:truncated-results}, ordered left to right. These values are for the full $12$ parameter fit, i.e., those including $\trad$ as a parameter. $\chi^2$ is calculated in a mask surrounding the images (see \citetalias{ORiordan2019} \S 3.3). The number of degrees of freedom, $N_\mathrm{dof}$, is the number of pixels in the mask minus the number of fitted parameters}

\end{table}

The results of fitting to the mock observations are provided in \cref{fig:truncated-results}, plotting posterior probability contours in a corner plot of the  interesting parameters, i.e. $t$ and $q$ for Case 1 (middle row), and $t$, $q$ and $\trad$ for Case 2 (bottom row). To check the goodness of fit of the recovered parameter values we compare the mock observations with the best fit model, made using the median sample values from the posteriors in \cref{fig:truncated-results}. For a large number of degrees of freedom $N_\mathrm{dof}$, $\chi^2$ is normally distributed with mean $N_\mathrm{dof}$ and standard deviation $\sqrt{2N_\mathrm{dof}}$. Therefore, if the model is a good fit to the data, one expects the quantity $\left(\chi^2-N_\mathrm{dof}\right)/\sqrt{2N_\mathrm{dof}}$ to lie within $\pm2$ of zero $95$\% of the time. This quantity is printed in \cref{table:chisq} and shows that the recovered parameter values are indeed a good fit to the mock observations.

In each case we see that the parameters are more precisely constrained as the axis ratio $q$ decreases, and ellipticity increases, left to right. This is as expected, considering that the circular case provides no constraints on the mass profile. For Case 1, with $\trad$ fixed, the expected anticorrelation between $q$ and $t$ is clearly seen i.e. as $t$ increases, and the potential becomes more circular, this is compensated for by a decrease in $q$, or increase in ellipticity. For Case 2, where $\trad$ is an extra free parameter, the main correlation is now between $q$ and $\trad$: as $\trad$ increases the potential becomes less circular, so $q$ must increase, i.e. ellipticity must decrease, to compensate. Interestingly the constraints on $t$ are very similar for the same $q$, between Cases 1 and 2. We also created images with different values of $\trad$ and confirmed that the constraints on $t$ improve as $\trad$ increases, as expected.

These results show that, although there is no mass in the annulus spanned by the images, the recovered values for the slope $t$ are consistent with the input values. It is clear that for a non-circular lens, the images do provide a constraint on the interior mass profile. This constraint is found to improve with ellipticity, and as the truncation radius approaches the radius of the images. In the best case constraints of $\sigma_t<0.05$ can be achieved on the interior slope. In \cref{sec:introduction} we drew attention to a belief frequently found in the literature that lensing measures the slope of the profile at the Einstein radius, and not interior to the images. The results presented here contradict this belief.

\section{Summary}
\label{sec:summary}

In this paper we have derived the lensing properties of the 2D broken power-law model, a versatile model for determining the projected mass profile of the lens, comprising power-laws of slope $t_1$ and $t_2$ either side of the break radius $\brad$. This model is the simplest extension of the power-law traditionally used in lensing studies, adding two degrees of freedom. We also presented a method for the efficient computation of the BPL model. 

The BPL model derives from the truncated power-law model. In the TPL model the surface density is described by a power-law inside some elliptical radius, called the truncation radius, and is zero outside. Using mock observations constructed with the TPL model we showed that, for an elliptical lens, the observations can constrain the slope interior to the images. The ellipticity of the lens and the truncation radius are also measurable. The sensitivity to the slope interior to the images improves with increasing ellipticity. This result contradicts the standard picture that lensing measures the slope of the mass profile at the Einstein radius and is insensitive to the mass profile interior to the images.

This demonstration opens up the possibility of using strong gravitational lensing to analyse the projected mass profile in detail. In the next paper we will present a full exploration of the usefulness of the BPL model for measuring mass profiles in lens galaxies. We will determine
the uncertainties on the parameters $t_1$, $t_2$, $\brad$ over the parameter space of source position and lens ellipticity. This identifies the particular configurations where all three parameters may be measured accurately, which are the most useful for detailed study, for example in measuring the Hubble constant.

\section*{Acknowledgements}
We are grateful to the Imperial College Research Computing Service for HPC resources and support. CMO'R is supported by a Science and Technology Facilities Council (STFC) Studentship.

\section*{Data Availability}
The data used in this paper are available from the corresponding author on request.

\bibliographystyle{mnras}
\bibliography{bibliography}

\begin{thebibliography}{}
\makeatletter
\relax
\def\mn@urlcharsother{\let\do\@makeother \do\$\do\&\do\#\do\^\do\_\do\%\do\~}
\def\mn@doi{\begingroup\mn@urlcharsother \@ifnextchar [ {\mn@doi@}
  {\mn@doi@[]}}
\def\mn@doi@[#1]#2{\def\@tempa{#1}\ifx\@tempa\@empty \href
  {http://dx.doi.org/#2} {doi:#2}\else \href {http://dx.doi.org/#2} {#1}\fi
  \endgroup}
\def\mn@eprint#1#2{\mn@eprint@#1:#2::\@nil}
\def\mn@eprint@arXiv#1{\href {http://arxiv.org/abs/#1} {{\tt arXiv:#1}}}
\def\mn@eprint@dblp#1{\href {http://dblp.uni-trier.de/rec/bibtex/#1.xml}
  {dblp:#1}}
\def\mn@eprint@#1:#2:#3:#4\@nil{\def\@tempa {#1}\def\@tempb {#2}\def\@tempc
  {#3}\ifx \@tempc \@empty \let \@tempc \@tempb \let \@tempb \@tempa \fi \ifx
  \@tempb \@empty \def\@tempb {arXiv}\fi \@ifundefined
  {mn@eprint@\@tempb}{\@tempb:\@tempc}{\expandafter \expandafter \csname
  mn@eprint@\@tempb\endcsname \expandafter{\@tempc}}}

\bibitem[\protect\citeauthoryear{{Bateman}}{{Bateman}}{1955}]{Bateman1995}
{Bateman} H.,  1955, {Higher transcendental functions}

\bibitem[\protect\citeauthoryear{{Bourassa} \& {Kantowski}}{{Bourassa} \&
  {Kantowski}}{1975}]{Bourassa1975}
{Bourassa} R.~R.,  {Kantowski} R.,  1975, \mn@doi [\apj] {10.1086/153300},
  \href {https://ui.adsabs.harvard.edu/abs/1975ApJ...195...13B} {195, 13}

\bibitem[\protect\citeauthoryear{{Chae}, {Bernardi}  \& {Kravtsov}}{{Chae}
  et~al.}{2014}]{Chae2014}
{Chae} K.-H.,  {Bernardi} M.,   {Kravtsov} A.~V.,  2014, \mn@doi [\mnras]
  {10.1093/mnras/stt2163}, \href
  {http://adsabs.harvard.edu/abs/2014MNRAS.437.3670C} {437, 3670}

\bibitem[\protect\citeauthoryear{{Du}, {Zhao}, {Fan}, {Shu}, {Li}  \&
  {Mao}}{{Du} et~al.}{2020}]{Du2020}
{Du} W.,  {Zhao} G.-B.,  {Fan} Z.,  {Shu} Y.,  {Li} R.,   {Mao} S.,  2020,
  \mn@doi [\apj] {10.3847/1538-4357/ab7a15}, \href
  {https://ui.adsabs.harvard.edu/abs/2020ApJ...892...62D} {892, 62}

\bibitem[\protect\citeauthoryear{{Hezaveh} et~al.,}{{Hezaveh}
  et~al.}{2016}]{Hezaveh2016}
{Hezaveh} Y.~D.,  et~al., 2016, \mn@doi [\apj] {10.3847/0004-637X/823/1/37},
  \href {http://adsabs.harvard.edu/abs/2016ApJ...823...37H} {823, 37}

\bibitem[\protect\citeauthoryear{{Kochanek}}{{Kochanek}}{1995}]{Kochanek1995}
{Kochanek} C.~S.,  1995, \mn@doi [\apj] {10.1086/175721}, \href
  {http://adsabs.harvard.edu/abs/1995ApJ...445..559K} {445, 559}

\bibitem[\protect\citeauthoryear{{Kochanek}}{{Kochanek}}{2006}]{Kochanek2006}
{Kochanek} C.~S.,  2006, in {Meylan} G.,  {Jetzer} P.,  {North} P.,
  {Schneider} P.,  {Kochanek} C.~S.,   {Wambsganss} J.,  eds, Saas-Fee Advanced
  Course 33: Gravitational Lensing: Strong, Weak and Micro. pp 91--268

\bibitem[\protect\citeauthoryear{{Kochanek}}{{Kochanek}}{2020}]{Kochanek2020}
{Kochanek} C.~S.,  2020, \mn@doi [\mnras] {10.1093/mnras/staa344}, \href
  {https://ui.adsabs.harvard.edu/abs/2020MNRAS.493.1725K} {493, 1725}

\bibitem[\protect\citeauthoryear{{Koopmans}, {Treu}, {Bolton}, {Burles}  \&
  {Moustakas}}{{Koopmans} et~al.}{2006}]{Koopmans2006}
{Koopmans} L.~V.~E.,  {Treu} T.,  {Bolton} A.~S.,  {Burles} S.,   {Moustakas}
  L.~A.,  2006, \mn@doi [\apj] {10.1086/505696}, \href
  {http://adsabs.harvard.edu/abs/2006ApJ...649..599K} {649, 599}

\bibitem[\protect\citeauthoryear{{Nightingale}, {Dye}  \&
  {Massey}}{{Nightingale} et~al.}{2018}]{Nightingale2018}
{Nightingale} J.~W.,  {Dye} S.,   {Massey} R.~J.,  2018, \mn@doi [\mnras]
  {10.1093/mnras/sty1264}, \href
  {https://ui.adsabs.harvard.edu/abs/2018MNRAS.478.4738N} {478, 4738}

\bibitem[\protect\citeauthoryear{{O'Riordan}, {Warren}  \&
  {Mortlock}}{{O'Riordan} et~al.}{2019}]{ORiordan2019}
{O'Riordan} C.~M.,  {Warren} S.~J.,   {Mortlock} D.~J.,  2019, \mn@doi [\mnras]
  {10.1093/mnras/stz1603}, \href
  {https://ui.adsabs.harvard.edu/abs/2019MNRAS.487.5143O} {487, 5143}

\bibitem[\protect\citeauthoryear{{O'Riordan}, {Warren}  \&
  {Mortlock}}{{O'Riordan} et~al.}{2020}]{ORiordan2020}
{O'Riordan} C.~M.,  {Warren} S.~J.,   {Mortlock} D.~J.,  2020, \mn@doi [\mnras]
  {10.1093/mnras/staa1697}, \href
  {https://ui.adsabs.harvard.edu/abs/2020MNRAS.496.3424O} {496, 3424}

\bibitem[\protect\citeauthoryear{{Schramm}}{{Schramm}}{1990}]{Schramm1990}
{Schramm} T.,  1990, \aap, \href
  {https://ui.adsabs.harvard.edu/abs/1990A&A...231...19S} {231, 19}

\bibitem[\protect\citeauthoryear{{Sonnenfeld}}{{Sonnenfeld}}{2018}]{Sonnenfeld2018}
{Sonnenfeld} A.,  2018, \mn@doi [\mnras] {10.1093/mnras/stx3105}, \href
  {http://adsabs.harvard.edu/abs/2018MNRAS.474.4648S} {474, 4648}

\bibitem[\protect\citeauthoryear{{Spingola}, {McKean}, {Auger}, {Fassnacht},
  {Koopmans}, {Lagattuta}  \& {Vegetti}}{{Spingola}
  et~al.}{2018}]{Spingola2018}
{Spingola} C.,  {McKean} J.~P.,  {Auger} M.~W.,  {Fassnacht} C.~D.,  {Koopmans}
  L.~V.~E.,  {Lagattuta} D.~J.,   {Vegetti} S.,  2018, \mn@doi [\mnras]
  {10.1093/mnras/sty1326}, \href
  {http://adsabs.harvard.edu/abs/2018MNRAS.478.4816S} {478, 4816}

\bibitem[\protect\citeauthoryear{{Suyu} et~al.,}{{Suyu}
  et~al.}{2017}]{Suyu2017}
{Suyu} S.~H.,  et~al., 2017, \mn@doi [\mnras] {10.1093/mnras/stx483}, \href
  {http://adsabs.harvard.edu/abs/2017MNRAS.468.2590S} {468, 2590}

\bibitem[\protect\citeauthoryear{Tessore \& Metcalf}{Tessore \&
  Metcalf}{2015}]{Tessore2015a}
Tessore N.,  Metcalf R.~B.,  2015, \mn@doi [\aap]
  {10.1051/0004-6361/201526773}, 580, A79

\bibitem[\protect\citeauthoryear{{Treu}}{{Treu}}{2010}]{Treu2010a}
{Treu} T.,  2010, \mn@doi [\araa] {10.1146/annurev-astro-081309-130924}, \href
  {http://adsabs.harvard.edu/abs/2010ARA%26A..48...87T} {48, 87}

\bibitem[\protect\citeauthoryear{{Wong} et~al.,}{{Wong}
  et~al.}{2020}]{Wong2020}
{Wong} K.~C.,  et~al., 2020, \mn@doi [\mnras] {10.1093/mnras/stz3094}, \href
  {https://ui.adsabs.harvard.edu/abs/2020MNRAS.tmp.1661W} {}

\bibitem[\protect\citeauthoryear{{Xu}, {Sluse}, {Schneider}, {Springel},
  {Vogelsberger}, {Nelson}  \& {Hernquist}}{{Xu} et~al.}{2016}]{Xu2016}
{Xu} D.,  {Sluse} D.,  {Schneider} P.,  {Springel} V.,  {Vogelsberger} M.,
  {Nelson} D.,   {Hernquist} L.,  2016, \mn@doi [\mnras]
  {10.1093/mnras/stv2708}, \href
  {http://adsabs.harvard.edu/abs/2016MNRAS.456..739X} {456, 739}

\bibitem[\protect\citeauthoryear{{van de Ven}, {Mandelbaum}  \& {Keeton}}{{van
  de Ven} et~al.}{2009}]{vandeVen2009}
{van de Ven} G.,  {Mandelbaum} R.,   {Keeton} C.~R.,  2009, \mn@doi [\mnras]
  {10.1111/j.1365-2966.2009.15167.x}, \href
  {http://adsabs.harvard.edu/abs/2009MNRAS.398..607V} {398, 607}

\makeatother
\end{thebibliography}

\bsp	
\label{lastpage}
\end{document}